%% file: main.tex
\documentclass[acmsmall,screen]{acmart}
\usepackage{algpseudocode}

\AtBeginDocument{%
  \providecommand\BibTeX{{%
    \normalfont B\kern-0.5em{\scshape i\kern-0.25em b}\kern-0.8em\TeX}}}

\renewcommand\footnotetextcopyrightpermission[1]{} 
\renewcommand\footnotetextauthorsaddresses[1]{} 

\setcopyright{none}

\acmConference[Programming Languages and the Law (ProLaLa)]{Workshop on Programming Languages and the Law}{2024}{U.S.}
\acmBooktitle{3rd International Workshop on Programming Languages and the Law (ProLaLa'24), 2024, California, U.S.}

\usepackage{array}
\usepackage[caption=false,font=normalsize,labelfont=sf,textfont=sf]{subfig}
\usepackage{textcomp}
\usepackage{stfloats}
\usepackage{url}
\usepackage{verbatim}
\usepackage{graphicx}
\def\BibTeX{{\rm B\kern-.05em{\sc i\kern-.025em b}\kern-.08em
    T\kern-.1667em\lower.7ex\hbox{E}\kern-.125emX}}
\usepackage{balance}

\usepackage{caption}
\usepackage{wrapfig}
\usepackage{tabu}

\input{header}

\begin{document}

\title[Metamorphic Debugging for Accountable Software]{Metamorphic Debugging for Accountable Software}

\author{Saeid Tizpaz-Niari}
\orcid{0000-0002-1375-3154}
\affiliation{%
  \institution{University of Texas at El Paso}
  \city{El Paso}
  \country{USA}
}
\email{saeid@utep.edu}

\author{Shiva Darian}
\orcid{0000-0002-0015-2771}
\affiliation{%
  \institution{New Mexico State University}
  \city{Las Cruces}
  \country{USA}
}
\email{shiva@nmsu.edu}

\author{Ashutosh Trivedi}
\orcid{0000-0001-9346-0126}
\affiliation{%
  \institution{University of Colorado Boulder}
  \city{Boulder}
  \country{USA}
}
\email{ashutosh.trivedi@colorado.edu}

\maketitle

\input{Sections/abstract_long}

\input{Sections/introduction}

\input{Sections/ReifyingAccountability}

\input{Sections/VerifyingAccountability}

\input{Sections/related}

\input{Sections/challenges-opportunities}

\input{Sections/conclusion}

\bibliographystyle{ACM-Reference-Format}
\bibliography{reference}

\end{document}

%% file: header.tex
\usepackage{hyperref}
\usepackage{multirow}

\newcommand{\pP}{\mathbb{P}}

\newcommand{\Dd}{\mathcal{D}}
\newcommand{\Hh}{\mathcal{H}}

\usepackage{xcolor}

\newcommand\vd[1]{d_{p}}

\newcommand{\set}[1]{\left\{ #1 \right\}}

\newcommand{\Real}{\mathbb R}

\newcommand{\Bool}{\mathbb B}

\usepackage[framemethod=tikz]{mdframed}
\usepackage[many]{tcolorbox}

\usepackage{framed}

 {\endMakeFramed}
 \definecolor{shadecolor}{gray}{0.9}

\definecolor{bgblue}{RGB}{245,243,253}
\definecolor{ttblue}{RGB}{91,194,224}

\mdfdefinestyle{mystyle}{%
  rightline=true,
  innerleftmargin=10,
  innerrightmargin=10,
  outerlinewidth=3pt,
  topline=false,
  rightline=false,
  bottomline=false,
  skipabove=\topsep,
  skipbelow=false
}

\newtcolorbox{myboxi}[1][]{
  breakable,
  title=#1,
  colback=white,
  colbacktitle=white,
  coltitle=black,
  fonttitle=\bfseries,
  bottomrule=0pt,
  toprule=0pt,
  leftrule=3pt,
  rightrule=3pt,
  titlerule=0pt,
  arc=0pt,
  outer arc=0pt,
  colframe=black!50,
}

\newtcolorbox{myboxii}[1][style=mystyle]{
  breakable,
  freelance,
  colback=white,
  colbacktitle=white,
  coltitle=black,
  fonttitle=\bfseries,
  bottomrule=0pt,
  boxrule=0pt,
  colframe=white,
  after skip=0pt,
  overlay unbroken and first={
    \draw[white!75!black,line width=3pt]
    ([yshift=-9pt]frame.north west) --
    ([yshift=9pt]frame.south west);
  },
  }

\newtcolorbox{myboxiii}[1][style=mystyle]{
  breakable,
  freelance,
  colback=white,
  colbacktitle=white,
  coltitle=black,
  fonttitle=\bfseries,
  bottomrule=0pt,
  boxrule=0pt,
  colframe=white,
  after skip=0pt,
  overlay unbroken and first={
    \draw[red!75!black,line width=3pt]
    ([yshift=-9pt]frame.north west) --
    ([yshift=9pt]frame.south west);
  },
}
\newtcolorbox{myboxv}[1][style=mystyle]{
  breakable,
  freelance,
  colback=white,
  colbacktitle=white,
  coltitle=black,
  fonttitle=\bfseries,
  bottomrule=0pt,
  boxrule=0pt,
  colframe=white,
  after skip=0pt,
  overlay unbroken and first={
    \draw[green!75!black,line width=3pt]
    ([yshift=-9pt]frame.north west) --
    ([yshift=9pt]frame.south west);
  },
  }

%% file: Sections/abstract_long.tex
\section*{Extended Abstract}
\label{sec:intro}
The availability of big data, combined with rapid progress in training AI agents, has enabled us to deploy software solutions that effectively address the increasing complexity and continual evolution of socio-economic and legal landscapes. 
Consequently, modern society is becoming increasingly reliant on software to resolve issues with legal, social, and economic implications. 
A case in point is tax preparation software, which now assists over 72 million Americans annually in filing their returns, significantly easing the collective burden of navigating the ever-growing complexity of U.S. tax laws.
Given their socio-economic and legally critical implications, \textit{ensuring software accountability---encompassing qualities such as legal compliance, explainability, perceptions of procedural justice, fairness of outcomes, and confidentiality/privacy---is of paramount social importance.}
Moreover, software that accurately interprets its requirements, complies with legal standards, and upholds social fairness can serve as a surrogate for legal and social norms, enabling policymakers to inquire about the law as seamlessly as a software engineer conducts a test.

Over forty years of research within the formal methods and software engineering communities have produced a range of principled approaches and powerful tools---such as formal property testing, automated fuzzing and debugging, program analysis, and satisfiability/constraint solving---that aid developers in enhancing the trustworthiness of critical software. 
\emph{Can we leverage this infrastructure to develop a computational framework that uncovers and explains accountability bugs in critical socio-legal software?}
Unfortunately, finding and explaining these accountability bugs in software implementations is remarkably difficult. 
A key obstacle stems from the challenge of reifying actual requirements in a formal, unambiguous language. The second challenge is the so-called \emph{oracle problem}, where the ideal expected output for a given input is unavailable, often requiring potentially adversarial deliberation by experts. 
Finally, due to evident privacy and legal concerns, trustworthy datasets are difficult to access.

Drawing from our experience in debugging U.S. tax preparation software, we propose that these challenges can be tackled by focusing on relational specifications.
While the exact output for a given input may be unknown, the relationship between the outputs of two related inputs may be easier to express. For example, to establish a higher tax benefit for blind individuals without explicit tax return oracles, one could compare two individuals differing only in their disability status, requiring the software to allocate higher tax benefits to the blind individual. This observation resembles i) the legal doctrine of precedent, or stare decisis (to stand by things decided), meaning that similar cases must yield similar rulings; and ii) metamorphic relations (MR) in software engineering that requires a specific relation between software inputs and outputs (e.g., perturbing \textit{race} attribute should not change the outcome of a tax return audit). \textit{We propose metamorphic debugging as the foundation for detecting, explaining, and repairing socio-legal software with respect to these relations.}

We demonstrate the applicability of metamorphic debugging over an open-source US tax preparation and a poverty management software. 
We show that challenging requirements from various US IRS  publications---including Form 1040, Publication 596, Schedule 8812, and Form 8863---as well as 
Pennsylvania benefits eligibility handbooks can be expressed as metamorphic relations. 
We also report critical accountability bugs found by two metamorphic debugging tools.

%% file: Sections/Introduction.tex
\section{Introduction}
\label{sec:intro}

As the laws and regulations have become more complicated and enormous, the role of software systems in navigating and understanding these intricacies has become more critical.
For example, tax preparation software has helped more than 72 million people in the United States to prepare their 2020 returns~\cite{IRS-rates}, significantly reducing the challenge of navigating the ever-increasing complexity of tax laws in the US. Similarly, legal practitioners and researchers have become considerably reliant on software code to implement models of legal code computationally. 

\begin{myboxi}
Given socio-economic and legal implications, it is critical to ensure software accountability---including properties such as \emph{legal compliance}, \emph{trust and procedural justice}, \emph{fairness}, and \emph{confidentiality/privacy}.
\end{myboxi}

Formal methods and software engineering communities have produced a range of principled approaches and practical tools---such as formal property testing, automated fuzzing and debugging, program analysis, and satisfiability/constraint solving---that help developers enhance the trustworthiness of critical software. 
\emph{Can we leverage this infrastructure to develop a computational framework that uncovers and explains accountability bugs in critical socio-legal software?}
However, finding and explaining accountability bugs in the implementation of social-legal critical software have remained a significant challenge due to factors such as:
\begin{enumerate}
    \item the difficulty of extracting formal specifications from the legal requirements, often expressed in the natural language;
    \item the unavailability of an oracle to resolve the ground truth for a query (oracle problem); and
    \item the lack of trustworthy datasets due to obvious privacy and legal concerns.
\end{enumerate}

We posit that since the US legal system adheres to \emph{stare decisis} (to stand by things decided) doctrine,
the specifications about the correctness of software can be written by comparing two or more similarly situated inputs as well as software. 
Such specifications are dubbed \emph{relational properties}. We consider a class of relational properties known as ``metamorphic relation'' (MR) that overcomes the oracle problem by comparing an input/software to its similar metamorphosed ones. For example, to establish a higher tax benefit for senior individuals without explicit tax return oracles, an MR could compare two individuals differing only in their age, requiring the software to allocate higher tax benefits to the senior individual
(e.g., age over 65). 
We put forward \emph{metamorphic debugging} as the foundation for detecting, explaining, and fixing sociolegal-critical software against metamorphic specifications to ensure accountability requirements.

\begin{myboxi}
Metamorphic debugging aims to accurately interpret accountability requirements, ensure compliance with legal standards, and certify software fairness. Thus, accountable software can serve as a surrogate for legal and social norms, enabling policymakers to inquire about the law as seamlessly as a software engineer conducts a test.
\end{myboxi}

In this work, we demonstrate the applicability of metamorphic debugging by focusing on open-source US tax preparation~\cite{ICSE-SEIS23} and multi-state poverty management software~\cite{escher2020exposing}. We show that challenging requirements from various US Internal Revenue Services (IRS) publications including Form 1040 (U.S. Individual Income Tax Return), Publication 596 (Earned Income Credit), Schedule 8812 (Qualifying Children and Other Dependents), and Form 8863 (Education Credits) as well as 
Pennsylvania benefits eligibility handbooks can be expressed as metamorphic relations. We also report critical 'accountability bugs' found by two recent works that adapted variants of metamorphic debugging approach in their works~\cite{ICSE-SEIS23,escher2020exposing}. 

\subsection{Case Study}

\noindent \textbf{Poverty Management System}.
In the United States, various assistance programs support low-income individuals, helping them access essential needs such as food, housing, healthcare, and other necessities.
Since application forms are complicated and require stringent demands for accuracy,
eligibility screening tools are offered online to help receive the benefits. 
Specifically, online benefits screening tools often advise households about their eligibility before proceeding with full applications. For example, the “Do I Qualify?” benefits screening tool provides eligibility predictions for five different public benefits: healthcare, food assistance, cash assistance, free or reduced-price school meals, and subsidized child care. It implements the requirements of the Pennsylvania benefits eligibility handbooks. Since an error in such a system can deprive qualified families from receiving the essential benefits they are entitled to, it is crucial to ensure software faithfully implements legal requirements.  

\vspace{0.5 em}
\noindent \textbf{Tax-Preparation Software}
The US tax preparation has grown into a $\$11.2$bn industry, with more than 72 million people preparing their taxes with software without the help of tax professionals, a 24\% increase from 2019~\cite{IRS-rates}. The ever-increasing complexity of  US income tax laws has rendered manual preparation of tax returns cumbersome and error-prone. According to the IRS, $90$ percent of tax filers filed taxes electronically in 2020~\cite{IRS-efile}. Thus, it is important that automated assistant tools (henceforth, \emph{socio-legal software}) provide accurate information to users, especially since individuals---not the software---are often held legally responsible for any software errors, and vulnerable
citizens might miss essential credits, benefits, and deductions\footnote{\emph{{\bf Langley v. Comm’r, T.C. Memo. 2013-22.} The misuse of tax preparation software, even if unintentional or accidental, is no defense to accuracy-related penalties under section 6662.}}.

\subsection{Challenges and Approach} 

Over forty years of research within the formal methods and software engineering communities have produced an array of principled approaches and performant tools---such as formal property testing, automated fuzzing and debugging, program analysis, and satisfiability/constraint solving---to aid developers in improving the trustworthiness of critical software. 
We seek to leverage this infrastructure in developing a \emph{computational framework} to uncover and explain accountability ``bugs" in critical socio-legal software. However, there are some concrete obstacles to this framework.
\begin{enumerate}
    \item {\it Reifying Accountability.} 
    The notion of accountability for socio-legal software systems is hard to reify.
    We present a metamorphic specification to overcome the oracle problem.

    \item {\it Verifying Accountability.}  Assuming that one can specify the accountability requirements,
    the next set of challenges is computational. We present a set of automated search-based
    strategy to detect and explain the root cause of bugs in socio-legal software systems.
\end{enumerate}

\subsubsection{Reifying Accountability} 
Relational properties~\cite{benton2004simple,yang2007relational} establish correctness by relating multiple inputs/output through a first-order logic specification.
We consider two classes of metamorphic relations\footnote{As a quick reminder, metamorphic testing is a software testing paradigm that tackles the oracle problem by considering software properties where the correctness of the software on an input does not require knowing the ``ground truth'' for that input; rather, the correctness can be validated by comparing the output of software for that input with the output of a slightly \emph{metamorphosed} one.  
For instance, consider software implementing a search engine: while there is no way to verify that the results returned for a particular keyword are correct (oracle problem), it is reasonable to expect that a correct search engine should return fewer results for a more restricted keyword.}~\cite{chen-original}: 
i) \textit{input perturbations} that require the outputs to remain the same between two similar inputs, running on two variants of (functionally) equivalent software; and ii) \textit{input/output perturbations} that require a specific relation between outputs of an input and its perturbation.

\begin{myboxi}
For example, to ensure a higher standard deduction for an individual with a blind spouse without explicit tax return oracles, an MR could compare two individuals differing only in their spouses' disability status, requiring the software to allocate higher tax benefits to the individual with a blind spouse.   
\end{myboxi}

\vspace{0.5 em}
\noindent \textbf{Metamorphic Inputs.} 
Escher and Banovic~\cite{escher2020exposing} developed a framework to test online poverty screening tools as implemented in the Pennsylvania ``Do I Qualify?'' website~\cite{COMPASS-HHS}. They took the benefit eligibility handbook~\cite{Penn-Human} and implemented their own screening software (the ground truth). Then, they use census data to generate test households and compare the results of the online tool (a potentially faulty software) against their implementations. 
This is an instance of relational properties that find bugs by comparing two variants of the same software that may accept different inputs, which are semantically equivalent inputs. 

\vspace{0.5 em}
\noindent \textbf{Metamorphic Inputs/Outputs.} 
\textsc{TenForty} is a data-driven technique~\cite{ICSE-SEIS23} to reify and verify
the accountability of \textit{OpenTaxSolver}~\cite{openTaxSolver} (tax years of 2018, 2019, 2020, and 2021),
a popular open-source tax preparation software~\cite{reddit-opentaxsolver,opensource-opentaxsolver}, in the domains of disability, credits, and deductions that are known to be challenging and error-prone~\cite{IRS-common-mistakes}. 
A key observation made in this preliminary work is that several compliance specifications can be expressed relating an individual with a counterfactual one. Tizpaz-Niari et al.~\cite{ICSE-SEIS23} present a formal (first-order) logic (\emph{metamorphic relations}) to express such compliance properties.
In collaboration with legal and tax experts, the authors explicated metamorphic relations for a set of challenging properties from various US Internal Revenue Services (IRS) publications including Form 1040, Publication 596 (EITC), Schedule 8812 (Qualifying Dependents), and Form 8863 (Education Credits). 

\begin{figure}[t]
    \centering
    \includegraphics[width=1.0\textwidth]{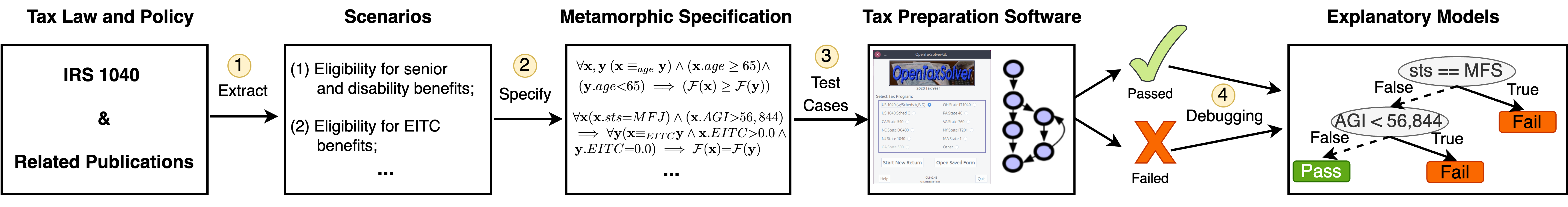}
    \caption{
    \textsc{TenForty}: General Framework using Disability and EITC benefits~\cite{ICSE-SEIS23}.
    }
    \label{fig:overall-framewok}
\end{figure}

\subsubsection{Verifying Accountability} 
Once notions of correctness are extracted and formalized, we use them to verify the correctness of concrete social-critical software.

\vspace{0.5 em}
\noindent \textbf{Metamorphic Verification.} 
\textsc{TenForty}~\cite{ICSE-SEIS23} is an open-source testing and debugging 
(Figure~\ref{fig:overall-framewok}), designed to test and debug tax software.
While it currently focuses on an open-source tax preparation software  \textit{OpenTaxSolver}~\cite{openTaxSolver} for the accompanied case study, it can be readily extended to other commercial software. 
{\textsc{TenForty}} generates tens of thousands of random test cases using a given compliance requirements as a metamorphic relation. Furthermore, it explains the circumstances under which the software has failed to comply using an explainable ML model (based on CART decision tree algorithm~\cite{Breiman/1984/CART}). 
{\textsc{TenForty}} has already revealed three types of failures in \textit{OpenTaxSolver}: missing some eligibility conditions (e.g., married people filing separately status is not eligible to take earned income credits); software fails to satisfy the correctness requirements when the computed tax returns get very close to zero (small non-zero values); and the updated software (e.g., 2021 version updated from 2020 version) that allows users to explicitly opt for an option that does not satisfy some correctness requirements in the corner cases. 
Similarly, Escher and Banovic~\cite{escher2020exposing} sampled random data points from the target populations of US Census Data and compared the outcome of their software to the online poverty screening tools (Pennsylvania ``Do I Qualify?'' website~\cite{COMPASS-HHS}). Any differences are deemed faulty. They found that for the subsidized
child care benefit, the ``Do I Qualify'' predicted every test household was ineligible, though at least 4.6\% of their test households were potentially eligible for the benefit.

%% file: Sections/ReifyingAccountability.tex
\section{Reifying Accountability}
\label{sec:reifying}
While various norms and exceptions are regularly updated by the US Internal Revenue Services (IRS) in various publications---such as Form 1040 (U.S. Individual Income Tax Return), Publication 596 (Earned Income Tax Credit), Schedule 8812 (Qualifying Children and Other Dependents), and Form 8863 (Education Credits)---these documents are expressed in natural language and may allow ambiguous interpretations. 
Moreover, such natural language specifications are not conducive to automatic software testing or analysis.

In the absence of explicit correctness requirements, pre-existing datasets can be used to test, debug, and quantify the belief in the correctness of the system. 
Unfortunately, it is difficult to obtain a meaningfully labeled dataset---individuals and their ``optimal'' tax returns---due to obvious privacy and legal concerns.
Even when one can learn a good generative model~\cite{mehta2022enhancement} to produce a synthetic population, tax software suffers from what is known as \emph{the oracle problem}~\cite{6963470} in software engineering: determining the correct output of an individual decision is time-consuming, expensive, and error-prone due to its highly subjective nature as discussed next.

\subsection{Metamorphic Specifications}
The first hurdle in developing a computational framework is to explicate an appropriate notion of correctness.
Note that, given the relevant information about an individual, resolving the correct tax decision for that individual requires accounting, legal, and ethical expertise.
Hence, obtaining an oracle for testing, debugging, and validation is impractically expensive.
\vspace{-0.1em}
\begin{myboxi}
The US tax law is principally derived from the \emph{common law} system and inherits the tenet of \textit{stare decisis} (i.e. the legal doctrine of precedent): similar cases must follow similar rulings. 
\end{myboxi}
As a corollary, the correctness of tax preparation software must also be viewed in comparison with similar cases. 
The correctness must be expressed as relations between two taxpayers with similar situations. 
Following the metamorphic testing, such properties are called \emph{metamorphic relations}. 

\vspace{0.5em}
\noindent \textbf{Formalizing MR for Input/Output Relation.}
For an individual ${\bf x}$, the label
${\bf x}.{\tt lab}$ shows the value of variable ${\tt lab}$. Example of variables
are ${\tt age}$ (numerical variable), ${\tt blind}$ (Boolean variable),
and ${\tt sts}$ (filing status with values). For labels $L \subseteq \mathcal{L}$ and inputs
${\bf x} \in \mathcal{D}$, ${\bf y} \in \mathcal{D}$, we say that ${\bf y}$ is
a metamorphose of ${\bf x}$ with the exceptions of labels $L$, and we write ${\bf x} \equiv_{L} {\bf y}$ if 
$\forall \ell \not \in L \text{ we have that } {\bf x}.\ell = {\bf y}.\ell$.
A metamorphic relation is a first-order logic formula with variables in $X$, constants
from domains in $\mathcal{D}$, relation $\equiv_L$, comparisons $\set{<, \leq, = , \geq, >}$
over numeric variables, predicate $\neg$ (negation) for Boolean valued labels, real-valued function
for federal tax return $\mathcal{F}: \Dd \to \Real$, Boolean connectives $\wedge$, $\lor$, $\neg$, $\implies$, $\iff$, and quantifiers $\exists x. \phi(x)$ and $\forall x. \phi(x)$ with natural interpretations. 
W.l.o.g., we assume that the formulas are given in the prenex normal form, i.e. a block of quantifiers followed by a quantifier-free formula. 

\vspace{0.5em}
\noindent \textbf{Examples.} 
Tizpaz-Niari et al.~\cite{ICSE-SEIS23} characterized $16$ metamorphic relations from $5$ domains of the U.S. Individual Income Tax Return: (1) Credit for the Elderly or the Disabled~\cite{pub524}, a credit for taxpayers who are aged 65 or older or who are retired on permanent and total disability;
(2) Earned Income Tax Credit (EITC)~\cite{pub596}, a refundable tax credits for lower-income households;
(3) Child Tax Credit (CTC), a nonrefundable credit to reduce the taxes owed based on the number of qualifying children under the age of 17~\cite{8812};
(4) Educational Tax Credit (ETC) that helps students with the cost of higher education by lowering their owed taxes or increasing their refund~\cite{8863};
and (5) Itemized Deduction (ID) which is an option for taxpayers with significant tax deductible expenses~\cite{1040sa}. Recently, Srinivas et al.~\cite{srinivas2023potential} have extended this work and expressed dozens of complex statutory legal requirements in taxation via metamorphic relations. Some examples are given here:
\begin{itemize}
\item \textbf{Property 1:} An individual with the married filing jointly 
($MFJ$) status with a disabled spouse must 
receive similar or higher tax benefits than a similar individual without a disabled spouse.
\begin{flushleft}
$\forall {\bf x} ({\bf x}.sts = MFJ) \implies \forall {\bf y} (({\bf x} \equiv_{s\_blind} {\bf y}) \wedge 
({\bf x}.s\_blind \wedge \neg {\bf y}.s\_blind))
    \implies \mathcal{F}({\bf x}) \geq \mathcal{F}({\bf y})$
\end{flushleft}

\item \textbf{Property 2:} An individual with the married filing separately ($MFS$) status is ineligible for EITC.

\begin{flushleft}
$\forall {\bf x} ({\bf x}.sts = MFS) {\implies} {\forall} {\bf y} ({\bf x} {\equiv_{L27}} {\bf y} \wedge {\bf x}.L27 >0.0 \wedge {\bf y}.L27 = 0.0) {\implies} \mathcal{F}({\bf x}){=}\mathcal{F}({\bf y})$.    
\end{flushleft}

\item \textbf{Property 3:} An individual with the married filing jointly ($MFJ$) status with $AGI$ over $56,844$ is ineligible for EITC.

\begin{flushleft}
$\forall {\bf x} ({\bf x}.sts{=}MFJ) \wedge ({\bf x}.AGI{>}56,844) \implies \forall {\bf y} ({\bf x} {\equiv_{L27}} {\bf y} \wedge {\bf x}.L27{>}0.0 \wedge {\bf y}.L27{=}0.0){\implies}\mathcal{F}({\bf x}){=} \mathcal{F}({\bf y})$.    
\end{flushleft}

\item \textbf{Property 4:} An individual who qualifies for EITC (e.g., income below $56,844$) must receive a higher or equal return than a similar unqualified one.
\begin{flushleft}
$\forall {\bf x} ({\bf x}.sts{=}MFJ){\implies} \forall {\bf y} ({\bf x} {\equiv_{AGI}} {\bf y} \wedge {\bf x}.AGI{\leq}56,844  \wedge  {\bf y}.AGI{>}56,844) \lor  ({\bf x} {\equiv_{L27}} {\bf y} \wedge  {\bf x}.L27{>}0.0 \wedge 
{\bf y}.L27{=}0.0) \lor ({\bf x} {\equiv_{QC}} {\bf y} \wedge {\bf x}.QC{\geq}{\bf y}.QC){\implies}\mathcal{F}({\bf x}){\geq} \mathcal{F}({\bf y}).$
\end{flushleft}

\item 
\textbf{Property 5:} This 4-property requires a comparison between four
``similar'' individuals as the rule changes for individuals with $AGI$ below $160$k and between $160$k and $180$k.
By holding $AGI$ constant between two individuals with $AGI$ below $160$k (varying the $ETC$ claims) and 
two individuals with $AGI$ between $160$k and $180$k (varying
the $ETC$ claims with the same rate), the property requires that individuals with lower income (below $160k$) receive higher or similar
tax returns.
\begin{flushleft}
$\forall {\bf x}, {\bf x'} ({\bf x}.sts {=} {\bf x'}.sts {=} MFJ) 
 \wedge ({\bf x}.AGI {\leq} 160k) \wedge (160k{<}{\bf x'}.AGI {<} 180k) 
 \implies \forall {\bf y}, {\bf y'} (({\bf x} \equiv_{L29} {\bf y})  \wedge
  ({\bf x'} \equiv_{L29} {\bf y'}) \wedge ({\bf x}.L29={\bf x'}.L29 \geq {\bf y}.L29={\bf y'}.L29)) 
 {\implies} ({\mathcal{F}}({\bf x}){-}  {\mathcal{F}}({y})){\geq}({\mathcal{F}}(x'){-} {\mathcal{F}}(y'))$  
\end{flushleft}

\end{itemize}

\vspace{0.5em}
\noindent\textit{Metamorphic Relations in Different Years.} 
As the tax law has evolved over different years,
the metamorphic relations need to be updated to reflect those changes.
Within the specified $16$ metamorphic relation~\cite{ICSE-SEIS23}, they identified changes in tax years 2018, 2019, and 2021 concerning the tax year 2020. For example, the $AGI$ requirements of $EITC$ differ over different years. Thus, they updated the metamorphic specification to ${\bf x}.AGI{>}54884$, ${\bf x}.AGI{>}55952$, and ${\bf x}.AGI{>}57414$ for tax years 2018, 2019, and 2021, respectively. Concretely, for tax year 2021, we can write:
\begin{flushleft}
$\forall {\bf x} ({\bf x}.sts{=}MFJ){\implies} \forall {\bf y} ({\bf x} {\equiv_{AGI}} {\bf y} \wedge {\bf x}.AGI{\leq}57,414  \wedge  {\bf y}.AGI{>}57,414) \lor  ({\bf x} {\equiv_{L27}} {\bf y} \wedge  {\bf x}.L27{>}0.0 \wedge 
{\bf y}.L27{=}0.0) \lor ({\bf x} {\equiv_{QC}} {\bf y} \wedge {\bf x}.QC{\geq}{\bf y}.QC){\implies}\mathcal{F}({\bf x}){\geq} \mathcal{F}({\bf y}).$
\end{flushleft}

\vspace{0.5 em}
\noindent \textbf{Formalizing MR with Input Relation.}
One class of reifying accountability detects bugs by ensuring two variants
of software (e.g., a potentially faulty software vs. a ground truth one) that should outcome the same results for two similar inputs (e.g., syntactically different, but semantically equivalent inputs). 
For an individual ${\bf x} \in \mathcal{D}$, the label
${\bf x}.{\tt lab}$ shows the value of variable ${\tt lab}$ where the domain $\mathcal{D}$ shows the (unknown) underlying distribution of population. Example of variables
are ${\tt size of household}$ (numerical variable), ${\tt blind}$ (Boolean variable),
and ${\tt num. children}$ (numerical variable). Let the ground truth software be $\mathcal{G}: \Dd \to \Bool$ and the target screening software be $\mathcal{F}: \Dd \to \Bool$. Then, a differential property simply requires that $\forall x,x'{\in}\mathcal{D}, x \equiv_{\set{}} x'{\implies} \mathcal{G}(x){==}\mathcal{F}(x')$. 

\vspace{0.5em}
\noindent \textbf{Examples.} Let us consider an example of the subsidized childcare benefit, the "Do I Qualify" predicted every test household was ineligible, though at least 4.6\% of their test households were potentially eligible for the benefit. Then, we can say $\exists x \in \mathcal{D}, {\bf x}.{\tt num. children} >= 1 \land {\bf x}\equiv_{\set{}}{\bf x}' \implies \mathcal{G}(x) \neq \mathcal{F}(x')$. Since they assume $\mathcal{F}$ is the ground truth; then any variations are deemed bugs.

%% file: Sections/VerifyingAccountability.tex
\section{Verifying Accountability}
For a given specification, searching the space of the inputs of socially-critical software is bound to be computationally demanding.
Even for data-driven (dynamic) analyses, detecting and explaining software inaccuracies involves comparing two similar inputs or software. 
Moreover, due to the aforementioned challenges, there is limited tool support in detecting, explaining, and repairing socially-critical software.

\subsection{Validation with Metamorphic Specification} 

\vspace{0.5 em}
\noindent \textbf{Statistical Validation of Tax Software.}
\textsc{TenForty}~\cite{ICSE-SEIS23} aims to falsify the correctness requirements by finding inputs that satisfy the pre-conditions of metamorphic relations, but do not satisfy the post-conditions.
In each step, the algorithm selects and perturbs inputs from the set of promising inputs to maximize the deviant from the expected outcome. If the deviant is more than a threshold, it adds the inputs with a `failed' label. Otherwise, it adds the inputs with a `passed' label. Since the absence of evidence for failures in the dynamic testing approach does not imply correctness, \textsc{TenForty} used hypothesis testing to provide statistical confidence on the
absence of failures, starting from the base test case. The null and alternative hypotheses are
$\Hh_0: \pP(x_1) \geq \theta,~~\Hh_1:\pP(x_1) < \theta$
where $\pP(x_1)$ is the probability that follow-up test cases, starting from $x_1$ as the source test,
are passed, $\theta$ is the lower-bound on the probability, $\Hh_0$ is null hypothesis, and $\Hh_1$ is the alternative. It follows Jeffreys test~\cite{jha2009bayesian,sankaranarayanan2013static}, a variant of Bayes factor,
with a uniform before finding a lower-bound on the number of
successive samples $K$ that is sufficient for us to convince $\Hh_0$:
$K \geq \lceil(-\log_2 B)/(\log_2 \theta)\rceil$
where $B$ in the numerator is the Bayes factor and can be set to $100$ for very strong evidence. 

\begin{figure}[!tb]
    \begin{minipage}[ht]{.47\linewidth}
    \centering
    \resizebox{\textwidth}{!}{%
    \begin{tabu}[ht]{|l|llll|llll|}
    \hline
    \multirow{2}{*}{\textbf{Property}} & \multicolumn{4}{c|}{\textbf{OpenTaxSolver 2020}} & \multicolumn{4}{c|}{\textbf{OpenTaxSolver 2021}} \\
     &  \#\textit{test cases} & \#\textit{fail} & \#\textit{pass} & \textit{T$_{F}$(s)} &  \#\textit{test cases} & \#\textit{fail} & \#\textit{pass} & \textit{T$_{F}$(s)} \\
    \hline
    
    P.1 & $35,790$ & $0$ & $35,790$ & $N/A$ & $32,355$ & $0$ & $32,355$ & $N/A$ \\ \hline
    P.2 & $19,936$ & $16,381$ & $3,555$ & $0.05$ & $32,343$ & $0$ & $32,343$ & $N/A$ \\ \hline
    P.3 & $18,258$ & $18,258$ & $0$ & $0.04$ & $16,556$ & $16,556$ & $0$ & $0.05$ \\ \hline
    P.4 & $36,450$ & $0$ & $36,450$ & $N/A$ & $32,883$ & $0$ & $32,883$ & $N/A$ \\ \hline
    P.5 & $18,015$ & $0$ & $18,015$ & $N/A$ & $16,346$ & $0$ & $16,346$ & $N/A$ \\
    \hline
    \end{tabu}%
    }
    \end{minipage}
    \hspace{0.2 em}
    \begin{minipage}[ht]{.25\linewidth}
        \centering
        \includegraphics[width=1.0\textwidth]{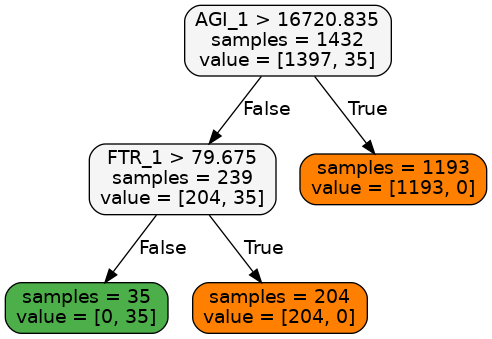}
    \end{minipage}
    \hspace{0.1 em}
    \begin{minipage}[ht]{.25\linewidth}
        \centering    
        \includegraphics[width=1.0\textwidth]{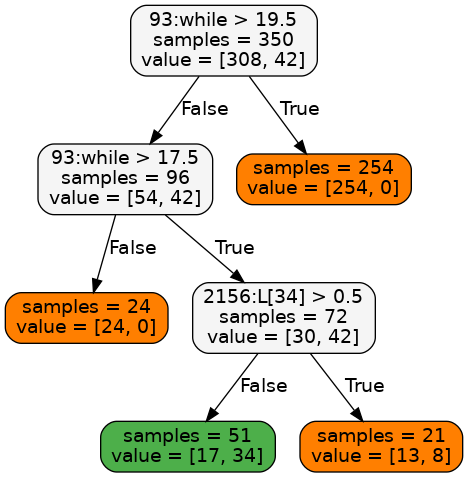}
    \end{minipage}
    \caption{(left) Metamorphic testing of $5$ metamorphic properties (see Section~\ref{sec:reifying}); (middle) Explanations in the input space. (Right) Explanations in the internal space.}
    \label{fig:enter-label}
\end{figure}

\textsc{TenForty}~\cite{ICSE-SEIS23} considered four versions of \textit{OpenTaxSolver}~\cite{openTaxSolver} (for years 2018, 2019, 2020, and 2021). 
The tool ran for $15$ mins over each version and its search strategy generated over $36,000$ test cases on average. For the preliminary metamorphic relations, there was no evidence of failures for tax years 2018 and 2019. However, there was evidence of failures for the 2020 and 2021 versions. One of these failures was related to an updated tax policy that set different income thresholds for eligibility for EITC!
Table~\ref{fig:enter-label} shows the results of test-case generations for properties 1-5. For example, OpenTaxSolver 2020 failed on MR (2) property, but it passed the equivalent requirement for 2021. Upon further investigation, we found that due to an update in the tax law (as published by the IRS), version 2021 turns out to be correct without any code update. 


\vspace{0.2 em}
\noindent \textbf{Data-Driven Debugging.}
\textsc{TenForty}~\cite{ICSE-SEIS23} explains the circumstances under which the software fails to
satisfy metamorphic requirements using CART decision tree inference.
\textsc{TenForty} provides debugging support in both
spaces of input variables (fields/ items in the tax forms) and internal variables (conditions, loops, and function calls of tax preparation software). The following decision trees (DT) show an explanation for the failure (orange and green leaf nodes correspond to failed passed test cases, respectively). The DT (left) shows that the software fails for all cases except when the AGI is below \$167k and the federal tax return is below \$79. The DT (right) shows the failure circumstances based on the internal properties of software: the number of loops taken in line 93 and the condition of $L$ variable in line 2,156 distinguish failed and passed test cases.

Escher and Banovic~\cite{escher2020exposing} sourced test households from a subset of Pennsylvania state data in the American Community Survey (ACS) Public Use Microdata Sample~\cite{USCB}. Then, they used Selenium, a browser automation tool, to enter their data entry into the screening tool (Pennsylvania ``Do I Qualify?'' website~\cite{COMPASS-HHS}) and retrieve the outcome via Beautiful Soup~\cite{Beautiful-Soup}. Finally, with their CS and Legal backgrounds, they compiled the eligibility guidelines for each benefit from the statute or policy manual and encoded them into a Python (ground truth) program that accepts a household as input and returns an eligibility prediction as output. For instance, they model the Supplemental Nutrition Assistance Program (SNAP)~\cite{SNAP-Handbook} benefit where the thirty-two chapters of the SNAP manual explain the food assistance requirements.
They report any discrepancy between the ground truth function and the Pennsylvania ``Do I Qualify?'' website~\cite{COMPASS-HHS} as a bug.

%% file: Sections/related.tex
\section{Related Work}
\label{sec:related}
\noindent {\bf Legal-Critical Software.} 
Let us first consider the work by Matthews et al.~\cite{10.1145/3306618.3314279,10.1145/3375627.3375807} on
forensic DNA software that aims to understand the role of black-box forensic
software in moral decision-making in criminal justice. They conduct independent
testing of Forensic Statistical Tool (FST), a forensic DNA system developed in
2010 by New York City’s Office of Chief Medical Examiner (OCME). Using a collection of over 400 mixed DNA samples, they found that an undisclosed data-dropping method in software impacts about 25\% of the samples and leads to shifted results
toward false inclusion of individuals who were not present in an evidence sample.
The techniques developed in our paper may allow one to uncover such vulnerabilities without having access to DNA samples purely based on an appropriate metamorphic query uncovering differential data-dropping.

\vspace{0.5 em}
\noindent {\bf Tax Prep Software.} 
Merigoux et al.~\cite{10.1145/3446804.3446850} developed a compiler for the French tax code. They pointed out multiple limitations of 
$M$ programming language, developed by the French Public Finances Directorate to write tax rules (available at~\cite{French-M-Rules}).
Then, they developed a domain-specific language that allows for specifying complicated rules and lifting them to modern languages like Python.
They also used a dynamic random search to validate the software, as opposed to formal verification, due to the large space of inputs, floating-point computations, and various optimizations. In comparison, the US has much less automation in tax preparation with no available specifications and regulated test cases. Yu, McCluskey, and Mukherjee~\cite{yu2020tax} proposed knowledge-based graphs to personalize tax preparation within the TurboTax software. In addition, SARA~\cite{DBLP:conf/kdd/HolzenbergerBD20} translated the statutes into Prolog programs and the cases into Prolog facts, such that each case can be answered by a single query. Alternatively, they proposed to adapt high-dimensional natural language processing techniques such as Legal BERT to overcome some of the intrinsic limitations of logic-based programming. 

\vspace{0.5 em}
\noindent {\bf Fairness.} 
Fairness has emerged as a critical safety requirement of AI-driven software
systems~\cite{tramer2017fairtest,galhotra2017fairness,10.1145/3377811.3380331,ICSE2022,DICE}.
Metamorphic testing has been significantly used to ensure the fairness of data-driven
and AI-driven systems~\cite{ma2020metamorphic,ribeiro-etal-2020-beyond,ma2020metamorphic}. 
Black et al.~\cite{10.1145/3531146.3533204} studied the fairness of algorithmic tax audit selections by the United States Internal Revenue Service (IRS) from 2010-14 using the concept of vertical equity. They found that flexible machine learning with higher accuracy, as opposed to simpler ones, may undermine vertical equity by shifting tax audit burdens from high-income to middle-income taxpayers. They also pointed out multiple weaknesses in applying existing fairness solutions to the tax audit problems. Finally, they made multiple suggestions to improve vertical equity through fair audit selection. Other articles also made similar observations that the IRS might audit low-income taxpayers at the same rate as the richest taxpayers~\cite{Bias-tax-notes,IRS-ProPublica-article,IRS-theatlantic-article}. For example, a ProPublica article reported that the top 1\% of taxpayers by income were audited at a rate of 1.56\% whereas earned income tax credit (EITC) recipients, who typically have an annual income under \$20,000, were audited at 1.41\% in 2018. Multiple works also pointed out the issue of the ``color-blind'' tax code in the US~\cite{Moran98,Bearer-Friend18,Bearer-Friend21}. Moran~\cite{Moran98}
used the horizontal equity as a fairness notion that requires similarly situated taxpayers should
be taxed similarly. The study found that black and white people who have similar
incomes are not taxed similarly because they fundamentally have different lifestyles. Studying fairness and ethical aspects of tax law as implemented in a tax preparation software system is an interesting future direction. 
Beyond the US tax system, a Dutch digital welfare fraud detection system, known as SyRI, is ruled to be unlawful since it does not comply with the
right to privacy under the European Convention of Human Rights and the European Union
General Data Protection Regulation (GDPR).
Furthermore, the court found that the SyRI legislation is insufficiently transparent
and verifiable and brings risks of discrimination~\cite{doi:10.1177/13882627211031257}.

%% file: Sections/challenges-opportunities.tex
\section{Future Directions}

We believe that there are significant opportunities to improve the accountability of social-critical software. Some challenges that require further study are: 

\begin{itemize}
    \item Role of large language models (LLMs) in extracting legal specification (NLP to FOL),
    \item The use of deep neural networks (DNNs) for approximating the legal-critical software,
    \item Using Generative Adversarial Networks (GANs) for learning data distributions/software,
    \item Accountable deployment of LLMs in the legal-critical setting with Human (expert) Feedback,
    \item A larger landscape of software testing and debugging.
\end{itemize}

\noindent Next, we outline a few exciting future directions. 

\subsection{Curating an Exhaustive Set of Metamorphic Requirements.} 
One direction is to extend the preliminary work to include a larger set of correctness requirements. Recently, Srinivas~\cite{srinivas2023potential} explored a few-shot in-context learning to generate metamorphic relations for tax preparation software. More work in this direction is needed for complicated forms like Form 1099-R (Distribution Form for Pensions, Annuities, Retirement, etc). In a particular scenario, when the form is checked with the distribution code of $7$ (normal distribution), the taxable amount of gross distribution is not included in Form 1099-R. In such a case, a worksheet called \texttt{simplified method} is used to compute the taxable amount (but only if the start date of the annuity is after July 2, 1986; otherwise, a more complicated method known as the generalized rule is used). The computation depends on multiple conditions such as the age at the annuity starting date, the annuity starting date, and the combined age of the primary with their beneficiary~\cite{Publication-721}. 
For example, take two similarly situated taxpayers between the ages of 66 and 70 with no beneficiaries who both have annuities. 
If Taxpayer A's annuity began several years before Taxpayer B's, Taxpayer A should receive higher tax benefits compared to Taxpayer B; however, if the annuities began at the same time, their tax returns should be similar. 
Considering the taxpayer's age and start date of annuity, we can develop $20$ metamorphic relations, including:
\begin{flushleft}
$\forall {\bf x}, {\bf y} \: ({\bf x} \equiv_{age,start} {\bf y}) \wedge (66 \leq {\bf x}.age,{\bf y}.age \leq 70) \wedge {\bf x}.start < 11.19.1996 \wedge {\bf y}.start > 11.18.1996 {\implies}(\mathcal{F}({\bf x}){\geq}\mathcal{F}({\bf y})).$
\end{flushleft}

\subsection{Causal Graph for Test-case Generation}
The causal dependencies between different fields/variables play a critical role to resolve the tax returns of individuals and generate test cases. 
Since there is a dearth of literature providing such models, in this part, one direction is to incorporate the experimental data as well as the tax-relevant data from the US Census to infer the causal dependencies between tax fields. The causal graph of dependencies between variables enables us to significantly improve the quality of our test cases where base and follow-up test cases can be computed over the causal graph via counterfactuals~\cite{pearl2009causality}. In addition, such models can detect any bugs in the design of tax policies that might lead to circular dependencies between variables. Moreover, it allows us to systematically measure the influence of one variable over all other variables via interventions.
Recent progress on causal testing~\cite{johnson2020causal,clark2023establishing,10.1145/3639478.3643530} shows significant potential in this direction. 


\subsection{Smart Fuzzing}
Automatic software testing has made substantial progress in recent years due to the advances in fuzzing and symbolic executions. Specifically, gray-box fuzzing has achieved tremendous success in detecting bugs and vulnerabilities
in critical software libraries. The evolutionary algorithm maintains a population of inputs
and mutates them randomly to generate new inputs~\cite{Klees2018EvaluateFuzzing}.
The key innovation in fuzzing is to use a \emph{lightweight} instrumentation that determines whether the current input is promising (e.g., by visiting a new edge in the control-flow graph)~\cite{AFL}. 
However, the current state-of-the-art often searches for general bugs and vulnerabilities such as memory leaks, out-of-bound access, and performance bugs. Establishing functional correctness of social-critical software requires techniques beyond the scope of current tools. 

\subsection{Data-driven Software Debugging}
The dependencies between variables
are very common in the tax code. For example, an individual with medical/dental
expenses ($MDE$) who files schedule $A$ is illegible for itemized deductions if
${\bf x}.$MDE$\leq{\bf x}.$AGI$*7.5\%$. If the software fails on this condition,
standard decision trees with box constraints will produce a large and uninterpretable
model. One idea is to exploit the simplicity of decision trees, but provide
explanations in richer domains such as Polyhedra
($\theta_1.{\bf x}.\ell_1 + \ldots + \theta_k*{\bf x}.\ell_k \leq d$). 
We hope that future work can infer fault-localization models in richer domains
(e.g., Polyhedra) for precise fault localization.

%% file: Sections/conclusion.tex
\section{Conclusion}
\label{sec:conclusion}
Modern society is becoming increasingly reliant on software solutions to resolve issues with significant legal, social, and economic implications.
Examples include criminal justice, tax auditing, and poverty management
systems. In this paper, we defined software accountability that includes properties
like legal compliance, explainability, perceptions of procedural justice, fairness of outcomes, and confidentiality/privacy. We emphasized the importance of accountable software to serve as a surrogate for legal and social norms, enabling policymakers to inquire about the law as seamlessly as a software engineer conducts a test.
We identified the challenges in ensuring accountability in software and demonstrated
the usefulness of metamorphic debugging to identify, explain, and repair accountability bugs in social- and legal-critical software. 
We showcased recent results that leverage metamorphic debugging to detect and explain accountability bugs in tax prep software and poverty management web systems.
We also outlined a set of promising future directions to build and verify accountability of sociolegal software. 

\vspace{0.25 em}
\noindent \textbf{Acknowledgement.} This research has been supported by NSF within the DASS program under grants CCF-2317206 and CCF-2317207. 